\documentclass[aps,showpacs,pra,twocolumn,superscriptaddress]{revtex4}
\usepackage{graphicx}
\usepackage{amssymb}
\usepackage{epstopdf}
\newcommand{\be}{\begin{equation}}
\newcommand{\ee}{  \end{equation}}
\newcommand{\ba}{\begin{eqnarray}}
\newcommand{\ea}{  \end{eqnarray}}
\newcommand{\ket}[1]{\left|#1\right>}
\newcommand{\bra}[1]{\left< #1 \right|}
\newcommand{\braket}[2]{\left< #1| #2 \right>}
\newcommand{\ketbra}[2]{\left| #1\right>\left< #2 \right|}

\begin{document}

\title{Geometric phase with nonunitary evolution 
in presence of a quantum critical bath}
\author{F. M. Cucchietti}
\affiliation{
ICFO -- Institut de Ci\`{e}ncies Fot\`{o}niques, Mediterranean Technology Park, 08860 Castelldefels, Spain}
\author{J.-F. Zhang}
\affiliation{Institute for Quantum Computing and Department of Physics,
University of Waterloo, Waterloo, Ontario, Canada N2L3G1 }
\author{F. C. Lombardo}
\affiliation{Departamento de F\'{\i}sica Juan Jos\'{e} Giambiagi, FCEyN UBA,
Facultad de Ciencias Exactas y Naturales, Ciudad Universitaria,
Pabell\'{o}n I, 1428 Buenos Aires, Argentina}
\author{P.I. Villar}
\affiliation{Departamento de F\'{\i}sica Juan Jos\'{e} Giambiagi, FCEyN UBA,
Facultad de Ciencias Exactas y Naturales, Ciudad Universitaria,
Pabell\'{o}n I, 1428 Buenos Aires, Argentina}
\affiliation{CASE Department,
Barcelona Supercomputing Center (BSC),29 Jordi Girona, 08034 Barcelona,Spain}
\author{R. Laflamme}
\affiliation{Institute for Quantum Computing and Department of Physics,
University of Waterloo, Waterloo, Ontario, Canada N2L3G1 }
\affiliation{Perimeter Institute for Theoretical Physics, Waterloo, Ontario, Canada N2J 2W9}

\date{\today}                                           

\begin{abstract}

Geometric phases, arising from cyclic evolutions in a curved parameter space,
appear in a wealth of physical settings. 
Recently, and largely motivated by the need of an experimentally 
realistic definition for quantum computing applications, 
the quantum geometric phase was generalized to open systems.
The definition takes a kinematical approach, with an initial state that is evolved cyclically but coupled to an environment
--- leading to a correction of the geometric phase 
with respect to the uncoupled case. 
We obtain this correction by measuring the nonunitary evolution of the reduced density matrix
of a spin one-half coupled to an environment.
In particular, we consider a bath that can be tuned near a quantum
phase transition, and demonstrate how the criticality information imprinted in the decoherence
factor translates into the geometric phase. 
The experiments are done with a NMR quantum simulator, in which the
critical environment is modeled using a one-qubit system.

\end{abstract}


\maketitle

For decades, the geometric phase \cite{berry} (GP) has fascinated physicists 
for its elegant theoretical grounds and its practical applications \cite{anandan}.
The GP is resilient to dynamical perturbations, thus, 
it might serve as a naturally fault-tolerant quantum information processing device \cite{cirac}.
In order to explore such applications, and unlike traditional studies of the GP
in closed systems with pure states, one must take into account realistic experimental
conditions --- i.e. the explicit presence of noise and environments.
Uhlmann was the first in considering 
a system in a mixed state, embeded, as a subsystem, in a larger system that is in a pure state \cite{uhlmann}. Later, 
Sj\"oqvist {\em et al} \cite{Sjoqvist} put forward 
a definition of
the GP for a general mixed state undergoing a cyclic unitary evolution ---
subsequently measured using NMR interferometry in Ref. \cite{Du}.
Different approaches to the problem were proposed \cite{openGP}. In the present Letter, we will follow 
the line of Tong {\em et al.} \cite{tong06}, who
developed a kinematic generalization of the GP to open systems that takes into account
the coupling to an environment (leading to a {\em nonunitary} evolution of 
the reduced density matrix of the system \cite{LombardoVillar}.) 
Arguably, this approach is better suited to explore the
usefulness of the GP in a real quantum computer undergoing decoherence processes \cite{SjoqvistPhysics}.
Here we report a measurement
of the GP for a  spin $1/2$ undergoing nonunitary evolution
induced by the coupling to an environment,
using the decoherence factor or fidelity decay \cite{Quan}.
In particular
---motivated by the recent observation
that baths near quantum criticality induce strong decoherence \cite{Quan}---
we choose an environment that can be tuned near a quantum phase transition (QPT).
This choice not only adds richness to the behavior of the GP, but also advances the program
of understanding it in general open systems
\cite{SjoqvistPhysics}.
In our experiments, performed in a NMR quantum simulator, 
we measure the full time dependence of the
decoherence factor of the system-spin --- from which we can determine the  
GP using the results of Ref. \cite{tong06}.
For the environment, we introduce a 
simple qualitative model of the ground state degeneracy that occurs at QPTs. 
%
Apart from demonstrating an alternative to traditional interferometry-based approaches 
for measuring the GP in open systems,
our results further establish the strong connections between quantum information,
quantum criticality, decoherence, and the quantum geometric phase 
\cite{Quan,GeometricTensor,GPcritical,carollo-pachos}
that have been the focus of much recent research
(especially in the context of quantum simulations \cite{Jingfu,JingfuFernando}).

The correction to the GP by a critical environment was first studied 
by Yi and Wang \cite{YiWang},
who gave some general analytical results and
found numerical instabilities in the GP of a qubit 
near criticallity of the bath (an $XY$ spin chain). 
More recently, it was shown that the GP of a spin coupled to an antiferromagnetic environment
changes suddenly when the bath undergoes a first order QPT \cite{Yuang}.
Notice that our problem is seemingly related to, but different than, 
the use of the GP as an order parameter in a QPT of a {\em closed}
system, as studied first by Carollo and Pachos and others \cite{carollo-pachos,GPcritical}.

We consider a spin $1/2$ coupled to an environment with a total Hamiltonian
$H=\Omega Z_{\cal S} \otimes I_{\cal E} + Z_{\cal S} \otimes H_{\cal S E} + I_{\cal S} \otimes H_{\cal E}$,
where $H_{\cal E}$ is the Hamiltonian of the bath, $Z_{\cal S}$ is the $z$ Pauli matrix of the system,
$I_{\cal S}$ is the identity operator of the system and $I_{\cal E}$ the one of the bath.
For simplicity, we only consider a dephasing spin--bath interaction,
$Z_{\cal S} \otimes H_{\cal S E}$, neglecting relaxation effects
and limiting the relevance of the initial state (see discussion below).
We take a product initial state for the spin-bath system,
$\rho(0)=\ketbra{\psi_0}{\psi_0}\otimes\ketbra{\varepsilon(0)}{\varepsilon(0)}$,
where $\ket{\psi_0}=\sin(\theta/2)\ket{0}+\cos(\theta/2)\ket{1}$ and $\ket{\varepsilon(0)}$ is a general
initial state of the bath.
In absence of the bath,
the spin follows an evolution around the Bloch sphere, reaching again the
initial state for $\tau=2\pi/\Omega$.
To compute the global phase gain during the evolution, one can use the Pancharatnam's definition \cite{Pancharatnam},
which contains a gauge dependent part (i.e a {\textit {dynamical}} phase $\Phi_d=-\pi \cos(\theta)$) and a
gauge independent part, commonly known as {\textit{geometric}} phase $\Phi_g=\pi(1- \cos(\theta))$.
When coupled to the bath, the reduced density matrix of the system at a time $t$ is
%
\ba 
&&\rho_{\rm r}(t)= \sin^2(\theta/2) \ketbra{0}{0} + \cos^2(\theta/2)
\ketbra{1}{1} \nonumber \\ && + \frac{\sin\theta}{2} e^{-i
2\Omega t}  r(t) \ketbra{0}{1} + \frac{\sin\theta}{2} e^{i
2\Omega t}  r^*(t)\ketbra{1}{0}. \label{rhor}
\ea 
Here, $r(t)=\braket{\varepsilon_0(t)}{\varepsilon_1(t)}$ is the decoherence factor induced 
by the environment, with
$\ket{\varepsilon_k(t)}=e^{-i t\left[ H_{B}+(-1)^k H_{\cal S E} \right]}
\ket{\varepsilon(0)}$.
The phase $\Phi $ acquired by the open system after a period $\tau$ is defined as
\cite{tong06}, 
\be \Phi = \arg \left[ \sum_{k}
\sqrt{\epsilon_k(\tau) \epsilon_k(0)} \braket{k(0)}{k(\tau)}
e^{-\int_0^\tau dt \bra{k(t)} \frac{\partial}{\partial t}\ket{k(t)}
} \right], \label{gp}\ee 
where $ \ket{k(t)}$ and $\epsilon_k(\tau)$
are respectively the instantaneous eigenvectors and eigenvalues of $\rho_{\rm r}(t)$. 
Of the two $k$ modes ($+$ and
$-$) of the one qubit model we are treating,
only the $+$ mode contributes to the GP.
Because
our environments can induce a complex decoherence factor,
i.e. $r(t)=|r(t)|e^{-i \varphi}$, 
we obtain a slightly different expression than that
shown in Ref. \cite{LombardoVillar}, namely
\ba
&\Phi& =  \int_0^\tau dt\left( \Omega
-\frac{\partial \varphi}{\partial t} \right)\sin^2(\frac{\theta^+_t}{2}) +  \nonumber \\
&\tan^{-1} &\frac{\sin \varphi(\tau)
\sin(\frac{\theta^+_\tau}{2})\sin(\frac{\theta}{2})}
 {\cos \varphi(\tau) \sin(\frac{\theta^+_\tau}{2})\sin(\frac{\theta}{2})
+\cos(\frac{\theta^+_\tau}{2})\cos(\frac{\theta}{2})},
\label{FinalGeometricPhase}
\ea
where we have defined
\ba
\cos(\theta^+_t/2)=\frac{2 (\epsilon_{+}-\sin^2(\theta/2))}{
\sqrt{|r(t)|^2 \sin^2(\theta)+4 (\epsilon_{+}-\sin^2(\theta/2))^2}} \\
\sin(\theta^+_t/2)=\frac{|r(t)| \sin(\theta)}{\sqrt{|r(t)|^2 \sin^2(\theta)+4 (\epsilon_{+}-\sin^2(\theta/2))^2}}.
\ea
During normal quantum evolution, the system gains a global phase.
The central result of Eq. (\ref{FinalGeometricPhase}) is to extract 
(by a proper choice of the ``parallel transport condition'') the purification
independent part of the phase --- which can be termed a geometric phase because it is gauge invariant
and reduces to the known results in the limit of a unitary evolution.

\begin{figure}[tbh]
\includegraphics[width=3.25in]{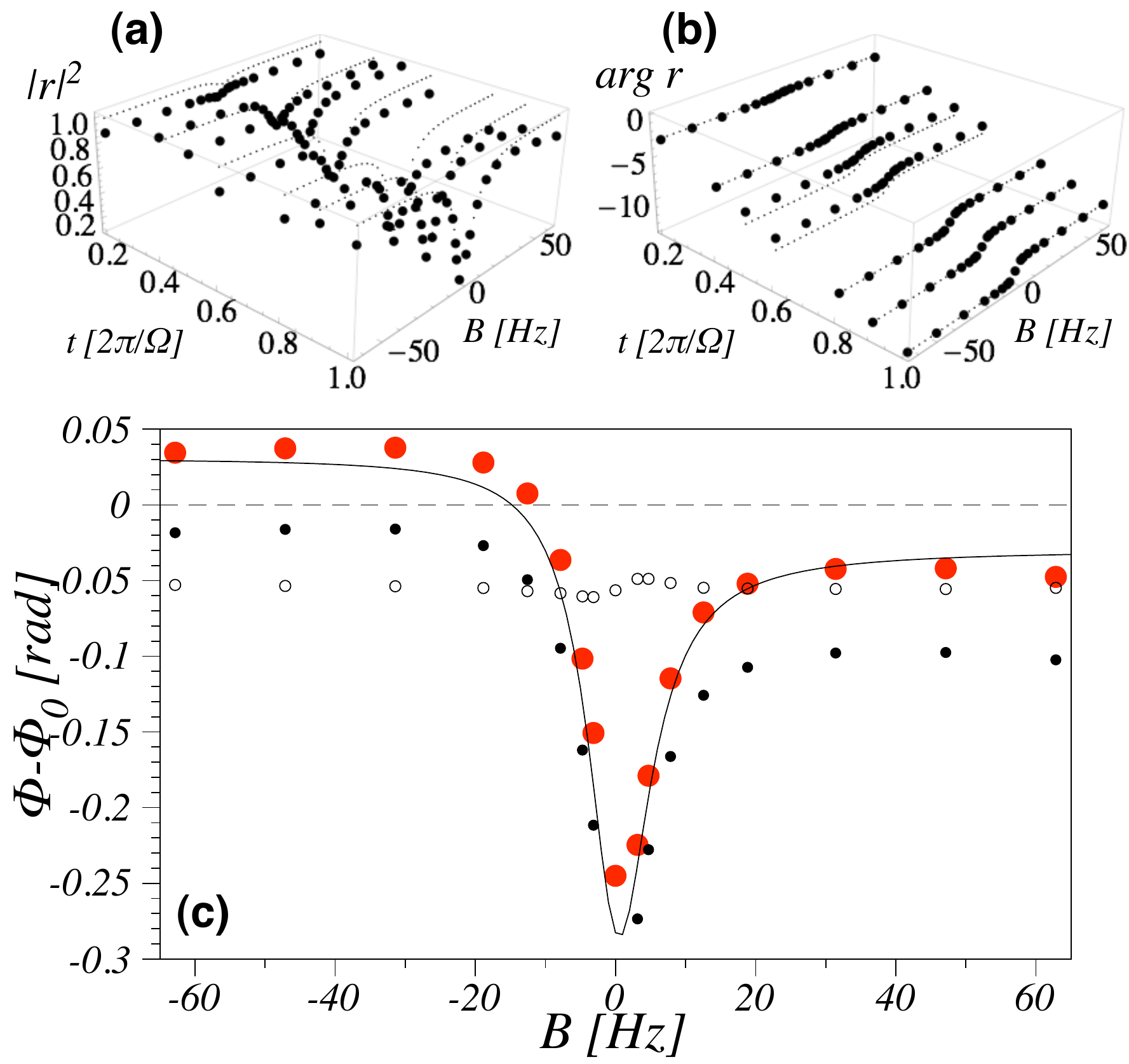}
\caption{
{\bf (a)} Observed absolute value squared of the decoherence factor and {\bf (b)} its argument,
both as a function of time and external magnetic field strength.
{\bf (c)} Computed correction to the geometric phase for a choice of $\theta=\pi/4$.
 In large filled circles is the experimental data, 
and in solid line is the theoretically expected value (without free parameters).
The corrected GP is the difference between the GP measured {\em in presence} of the
environment (small filled circles), and the GP measured when the system and environment are
{\em decoupled} (small empty circles).
In this setup, $\Omega=100\pi \ {\rm Hz}$, 
$\Delta=0.02\Omega$, and $\delta=0.1\Omega$. }
 \label{ExptFig}
\end{figure}

We have studied Eq. (\ref{FinalGeometricPhase}) 
both numerically and analytically 
with an Ising spin chain environment (see supplementary material \cite{EPAPS}), 
a paradigmatic example of a quantum phase transition.
In particular, the non-analiticity of the GP at the critical point in the 
thermodynamical limit becomes evident in the limit of weak system-bath coupling  \cite{EPAPS}.
Nevertheless, a full quantum simulation of a large enough critical system is 
on the edge of current technology, and beyond the scope of this Letter.
Therefore, 
before turning to the experimental results, let us discuss briefly our choice for modeling a critical 
environment.

Near its critical point, the
spectrum of a quantum critical system is characterized by the closing of the energy gap between the ground
and the first excited state. In the thermodynamical limit, the gap closes with a power law $\sim |\lambda - \lambda_c|^{z \nu}$
(where $z$ and $\nu$ are critical exponents), 
but for all finite size systems there is a minimum gap $\Delta$ near $\lambda_c$.
It is remarkable that, for many purposes,
this feature of the spectrum is enough to describe qualitatively
the effects of a critical environment:
as long as the excitations involved are small, and one is only interested in qualitative behavior,
a small energy expansion of the decoherence factor can
justify considering just two levels with appropriate dynamics \cite{JingfuFernando}.
Thus, we propose to mimic a complex critical bath 
using a simple two-level system
model with Hamiltonian $
H_{\cal E}=\lambda \left| \lambda \right|^{z \nu-1} \Delta \ Z_{\cal E} + \Delta \  X_{\cal E}$,
where $\lambda_c=0$ represents
the ``critical point'' or QPT.
The simplification might seem excessive, but it has been used successfully before:
For $z\nu=1$ (the mean field exponents),
it gives a correct qualitative description of
the creation of topological excitations during a finite speed quench \cite{damski}.
In essence, the model is quantitatively not far away from the small systems used in
demonstrations of quantum phase transitions \cite{ions,Jingfu}.
A complete characterization of when this model does not describe
the correct physics of a critical model is missing (one such example is the calculation of the path
length of an adiabatic evolution \cite{somma}).
Nevertheless, our results show that for the GP problem the model
gives a fair description when the gap $\Delta$ is much smaller than
the natural frequency $\Omega$ of the system spin.

Using a tomographic approach, we measure the GP of a  qubit coupled to a critical
environment using a nuclear magnetic resonance (NMR) quantum
simulator, with the environment represented by the two level model
described above (with critical exponents $z\nu=1$ and
a dimensionless transverse field strength $B=\lambda \Delta$). 
The target Hamiltonian to simulate experimentally is:
$H=\Omega Z_{\cal S} + \delta Z_{\cal S} Z_{\cal E} + B Z_{\cal E} + \Delta X_{\cal E}$,
where $Z_{\cal S}$ and $Z_{\cal E}$ are the $z$ Pauli matrices of the system and environment
respectively.
We obtain the GP by measuring the magnetization
of the system spin in the $X-Y$ plane, which gives us the decoherence factor $r(t)$.

Denoting by $\epsilon_\pm = \pm \Delta \sqrt{1+\lambda^{2z\nu}}$ the
eigenenergies of $H_{\cal E}$, the decoherence factor of this model is
\ba
r(t)&=& e^{i  \epsilon_-(\lambda)  t } \left[ \cos \epsilon_-(\lambda+\delta)  t  - \right. \nonumber \\
&i  & \left.   \frac{\epsilon^2_-(\lambda+\delta) -\Delta^2
\delta^2}{ \epsilon_-(\lambda) \epsilon_-(\lambda+\delta)  } \sin
\epsilon_-(\lambda+\delta)  t \right]. \ea 
where, to simplify notation, we have chosen the system--environment interaction to be
$H_{\cal S E}=\delta (I_{\cal S}-Z_{\cal S}) Z_{\cal E}$.
The correction to the GP 
due to this decoherence factor [shown in Fig.~(\ref{ExptFig}) with the experimental results
to be discussed below] contains the main elements observed in more complex models, as Ising spin 
chains \cite{EPAPS}:
a maximum correction of the GP at criticality, and a small asymmetric
correction far away from the critical point.
From this simplified model we can get insight into the physics of true quantum critical baths. 


Overall, the experimental sequence is as follows: We first fix
$\Omega$, $\delta$, and $\Delta$. Then, for each value of $B$, we
initialize the system, and measure the decoherence factor $r(t)$
of the system after evolution with an operator $U=e^{-i H t}$
for various times $t\in [0,2\pi/\Omega]$. 
The measured decoherence factor is shown in Fig.~(\ref{ExptFig}$a$)
and (\ref{ExptFig}$b$).
The GP is calculated using a numerical interpolation of $r(t)$
in Eq.~(\ref{FinalGeometricPhase}).

We choose the C$^{13}$ and H$^1$ spins in the molecule of carbon-13
labelled chloroform (CHCl$_3$) dissolved in d6-acetone as the
quantum registers (qubits) for the demonstration. The C$^{13}$ atom
simulates the system, and the
H$^1$  the environment,
where the scalar coupling between them is measured to be
$J=215$ Hz. Data were taken with a Bruker DRX 700 MHz spectrometer.

Our choice of system--environment coupling makes the decoherence factor $r(t)$ independent
of the initial state of the system (given by the angle $\theta$) [see Eq. (1)]. This, in turn, 
makes the GP depend trivially on $\theta$, which can be fully appreciated when approximating 
Eq. (\ref{FinalGeometricPhase}) in the weak coupling regime
(see supplementary material \cite{EPAPS}). Because we concentrate
on how the criticality properties of the bath affect the GP, it is experimentally convenient
to fix an initial state of the system that maximizes the signal to noise ratio, and 
change only the parameters of the environment spin. In particular, we choose the input
state of the system to be $(|0\rangle_{\cal S}+|1\rangle_{\cal S})/\sqrt{2}$. 
The corresponding decoherence factor can then be used to compute Eq. (\ref{FinalGeometricPhase}) 
for any other initial state of the system.
From  Eq. (1) we can see that $r(t)$ is encoded in the  coherent terms 
proportional to $\langle2\sigma_{+}\rangle$ [see Fig. (2$b$)], 
which  can be observed directly in NMR  by adding the two complex amplitudes of the peaks 
in the $C^{13}$ spectra.  

\begin{figure}[ht]
\includegraphics[width=3.25in]{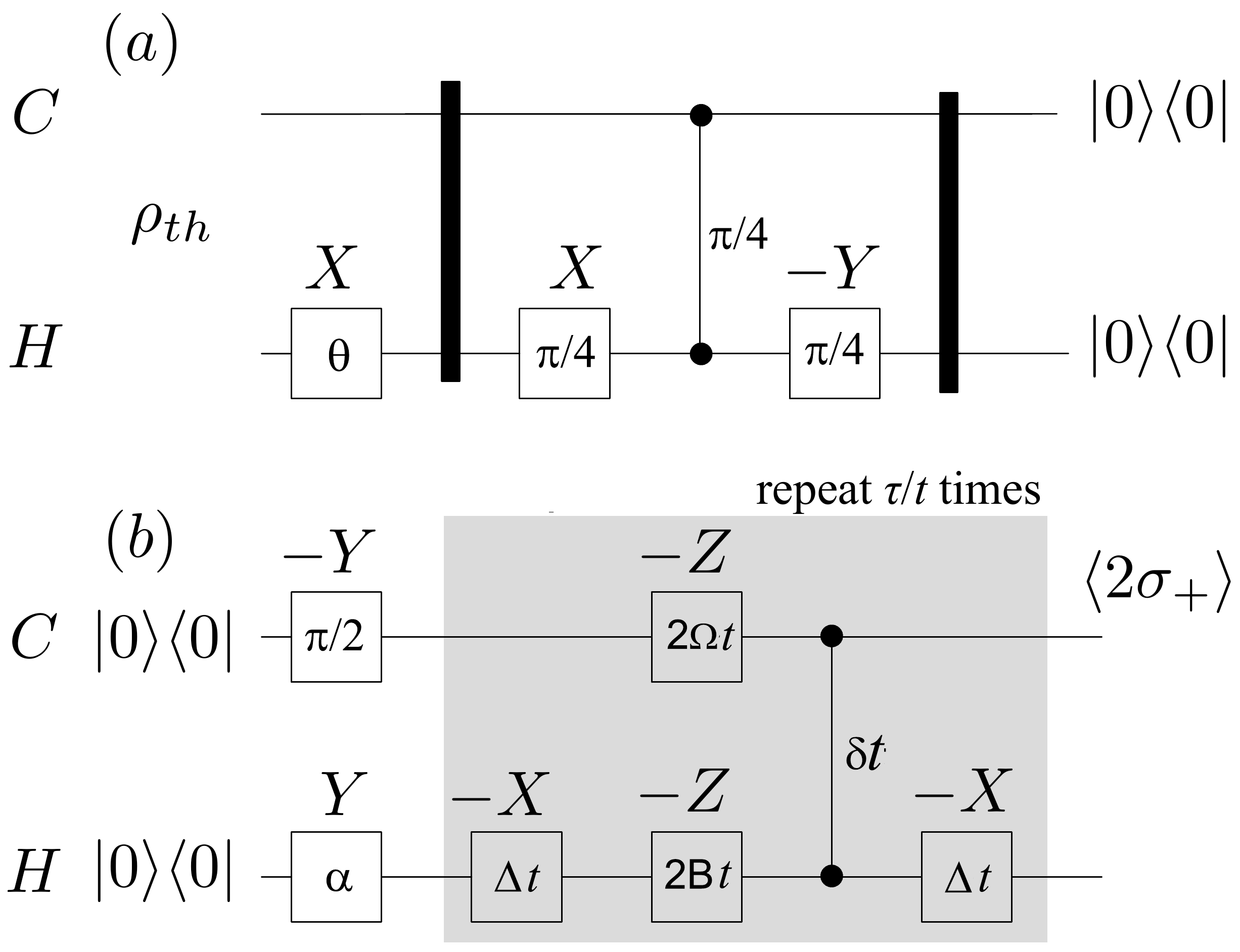}
\caption{$(a)$ Gate sequence for preparing the pseudo-pure state
$|00\rangle_{\cal S E}$ from the thermal state $\rho_{th} =\gamma_C
Z_{\cal S} + \gamma_H Z_{\cal E} $, where $\gamma_C$ and $\gamma_H$ denote the
gyromagnetic ratio of C$^{13}$ and H$^1$. 
 $(b)$ Gate sequence for the quantum simulation of the system and measurement of
 the decoherence factor
$r(\tau)$, which is proportional to $\langle2\sigma_{+}\rangle=\langle X+iY
\rangle$. 
In both plots the rectangles denote single-qubit gates,
implemented through radio-frequency pulses.
The rotation angle is shown inside the rectangle, and the direction above.
In the experiment we used $\cos\theta = 2\gamma_C/\gamma_H\approx 1/2$ and
$\tan\alpha=-\Delta/B$ with $\alpha\in(0,\pi)$.
The narrow black
rectangles denote the gradient pulses along $Z$-axis. The two filled
circles connected by a line denote the $J-$ coupling evolution
$e^{-i\phi Z_{\cal S}Z_{\cal E}}$, where $\phi$ is indicated next to the line.}
 \label{figpul}
\end{figure}

We use the gate sequence of  Fig.~(\ref{figpul}-a) \cite{pure1,pure2,pure3}
to prepare the pseudo-pure state $|00\rangle_{\cal S E}$, to which 
 we apply the unitary
 $e^{-i\pi Y_{\cal S}/4}e^{i\alpha Y_{\cal E}/2}$ to reach  the input state
$|\psi_{in}\rangle=(|0\rangle+|1\rangle)_{\cal S}|g\rangle_{\cal E}/\sqrt{2}$.
Here $|g\rangle_{\cal E}$ is the ground state of the environment
for a given $B$-value,
$|g\rangle_{\cal E} = |0\rangle\cos(\alpha/2)-|1\rangle\sin(\alpha/2)$,
where $\tan\alpha=-\Delta/B$. 
Because the decoherence factor is independent of the initial state of the system, 
we chose it such that it maximizes the signal-to-noise ratio of the experiment.

The quantum simulated evolution $U$ for a time $t$ can be implemented
with a Trotter approximation \cite{average1,average2},
\begin{equation}\label{Expevo}
    U \approx e^{-i\Delta t X_{\cal E}/2}
     e^{-i\delta t Z_{\cal S}Z_{\cal E}} 
     e^{-i\Omega t Z_{\cal S}} e^{-iB t Z_{\cal E}}
     e^{-i\Delta t X_{\cal E}/2}
\end{equation} 
where we choose $\Omega=100\pi \ {\rm Hz}$, 
$\Delta=0.02\Omega$, $\delta=0.1\Omega$,
and we apply the evolution operator $\tau/t$ times. 
We checked numerically that the Trotter approximation reduces the fidelity
less than $0.3\%$ for $B \in [-0.2\Omega,0.2\Omega]$.
Furthermore, we decompose the unitary operations $e^{-iB t Z_{\cal E}}$
as $e^{-i\pi X_{\cal E}/4} e^{-i B t Y_{\cal E}} e^{i\pi X_{\cal E}/4}$, 
 and 
 $e^{-i\Omega t Z_{\cal S}}$ 
 as $e^{-i\pi X_{\cal S}/4} e^{-i \Omega t Y_{\cal S}} e^{i\pi X_{\cal S}/4}$ 
 so that we can implement them
with standard  rf pulses. The coupling operation $e^{-i\delta t
Z_{\cal S}Z_{\cal E}}$ is realized using the natural spin coupling with an evolution
time $2\delta t/(\pi J)$.  After the evolution $U$, we measure
the magnetization in the $XY$ plane, which is proportional to the
decoherence factor $r(t)$. 
The whole gate sequence for
each measurement is shown in Fig. (\ref{figpul}-b). 
Notice that we measure the absolute
value as well as the complex phase of $r(t)$, necessary for the GP.
The total evolution time was always well below the natural decoherence time 
of the quantum simulator.

To eliminate systematic errors, we repeat the experiment but uncoupling the system
and the environment ($\delta=0$). From this we compute a baseline GP, which we subtract
from the full (coupled) experiment. Thus, we obtain 
the {\em correction} to the GP due to the presence of the critical environment, which
agrees well with theoretical expectations [see Fig. (\ref{ExptFig})$c$].


{\em Conclusions.} 
Using a NMR quantum simulator, 
we have obtained the quantum geometric phase of an open system undergoing nonunitary evolution.
The geometric phase is computed in a tomographic manner, i.e. we measure the off-diagonal elements of the
reduced density matrix of the system, from which we extract the decoherence
factor that we use in the definition of the open system GP.
In future work, we will introduce a third (probe) spin to perform an independent
and direct measurement of the GP using traditional interferometry-based 
techniques \cite{Du}.
Our experiments support the observation that when the environment is near a second order 
quantum phase transition, the correction to the GP becomes singular.
For our experiments, we introduced a simplified two-level
model that captures the essence of the spectral behavior of the critical environment:
the closing of the gap. 
%
By adding stochastic fields and further spins, we can
quantum-simulate more realistic environments and couplings to the system,
so that each initial state is affected differently by the bath.
Despite the apparent simplicity of our experiment, 
we believe that the techniques we developed are quite
general and applicable to more complex quantum simulations, and to 
related approaches such as bath engineering \cite{engbaths} --- designing an environment 
so that it induces a system to relax and decohere to interesting quantum many-body pure states.

We acknowledge helpful discussions with A. Ac\'{\i}n, J. I. Cirac, P. Hauke,
M. Lewenstein, G. Morigi, A. Roncaglia, and A. M. Souza. 
F.M.C. acknowledges financial support from Spanish MEC project TOQATA, ERC
Advanced Grant QUAGATUA, and Caixa Manresa. F.C.L is supported by
UBA, CONICET and ANPCyT--Argentina. P.I.V acknowledges financial
support from the UNESCO LOREAL Women in Science Programme.

\clearpage

\appendix

\section{Supplementary Material}

In these supplementary notes we show how the open system geometric phase (GP)
behaves in the limit of weak coupling to the environment. After analyzing the 
structure of the dominant terms, we will obtain analytically closed formulas for the
case of an Ising spin chain environment, and compare to exact numerical results. 
The analytical results will show explicitly the singularity of the GP
when the environment is at the critical point of a quantum phase transition.

\subsubsection{Small coupling expansion of the geometric phase}

The GP for an open system
[Eq. (3) in the main text] is
\ba
&\Phi& =  \int_0^\tau dt\left( \Omega
-\frac{\partial \varphi}{\partial t} \right)\sin^2(\frac{\theta^+_t}{2}) +  \nonumber \\
&\tan^{-1} &\frac{\sin \varphi(\tau)
\sin(\frac{\theta^+_\tau}{2})\sin(\frac{\theta}{2})}
 {\cos \varphi(\tau) \sin(\frac{\theta^+_\tau}{2})\sin(\frac{\theta}{2})
+\cos(\frac{\theta^+_\tau}{2})\cos(\frac{\theta}{2})},
\label{app:FinalGeometricPhase}
\ea
where 
\ba
\cos(\theta^+_t/2)=\frac{2 (\epsilon_{+}-\sin^2(\theta/2))}{
\sqrt{|r(t)|^2 \sin^2(\theta)+4 (\epsilon_{+}-\sin^2(\theta/2))^2}} \\
\sin(\theta^+_t/2)=\frac{|r(t)| \sin(\theta)}{\sqrt{|r(t)|^2 \sin^2(\theta)+4 (\epsilon_{+}-\sin^2(\theta/2))^2}},
\ea
and the only relevant eigenvalue of the reduced density matrix of the system is
\be
\varepsilon_+ =\frac{1}{2} \left( 1+ \sqrt{\cos^2(\theta)+ |r|^2 \sin^2(\theta)}
\right).
\ee
We want to obtain a more tractable expression of the GP in the limit of small coupling between the system and the
environment. For this, we expand the decoherence factor $r(t)=|r(t)| e^{i \varphi(t)}$ in powers of the
system--environment coupling strength $\delta$,
\ba
|r(t)|^2&=&1-R_{(2)}(t)\delta^2-R_{(3)}(t)\delta^3+{\cal O}(\delta^4) \\
\varphi(t) &=& \varphi_{(1)}(t)\delta+\varphi_{(2)}(t)\delta^2+\varphi_{(3)}(t)\delta^3 +{\cal O}(\delta^4) \nonumber
\ea
The zeroth order in $\varphi$ can be assimilated as an overall phase in the environment, and
the first order in $|r(t)|^2$ is zero for environments with a finite spectral band width, as with
spin environments \cite{app:CucchiettiPazZurek}. 
At this level of approximation, the arc-tangent term in Eq. (\ref{app:FinalGeometricPhase})
can be expanded as
\begin{widetext}
\ba
\sin^2\frac{\theta}{2} \left\{
\varphi_{(1)}(\tau) \delta +  \varphi_{(2)}(\tau) \delta^2 +
\left[ \varphi_{(3)}(\tau) - \frac{1}{2} \cos\theta \cos^2\frac{\theta}{2}
\varphi_{(1)}(\tau) \left( R_{(2)}(\tau)+\frac{\varphi^2_{(1)}(\tau)}{3} \right)
\right]
\right\},
\ea
while the term with the integral is

\ba
& \pi & (1-\cos\theta)- \sin^2\frac{\theta}{2} \varphi_{(1)}(\tau) \delta -
\left[
 \sin^2\frac{\theta}{2} \varphi_{(2)}(\tau) +
 \frac{\Omega}{4}\cos\theta\sin^2\theta \left( \int_0^\tau dt \ R_{(2)}(t)\right)
\right] \delta^2 \nonumber \\
&-& \left[ 
 \sin^2\frac{\theta}{2} \varphi_{(3)}(\tau) + \frac{1}{4} \cos\theta\sin^2\theta
 \left( \Omega
 \int_0^\tau dt \ R_{(3)}(t) -  \int_0^\tau dt \ R_{(2)}(t) \frac{\partial \varphi_{(1)}}{\partial t}(t)
 \right)
\right]\delta^3,
\ea
where we have assumed that $ \varphi_{(1)}(0)= \varphi_{(2)}(0)=\varphi_{(3)}(0)=0$.
Adding up the two contributions results in
\ba
\Phi& \simeq &
\pi  (1-\cos \theta )-
\cos \theta \sin^2 \theta 
\left[
\delta ^2 \frac{1}{4} \Omega \int_0^{\tau} R_{(2)}(t) \ dt  \nonumber \right. \\ & + & \left.
\frac{1}{24}   \delta ^3 \left(3 R_{(2)}(\tau)\varphi_{(1)}(\tau)  +\varphi^3_{(1)}(\tau) +
6 \Omega \int_0^{\tau}  R_{(3)}(t) \ dt  -6 \int_0^{\tau}  R_{(2)}(t) \frac{\partial \varphi_{(1)}}{\partial t}(t) \ dt   \right)
\right].
\label{app:ApproxGP} 
\ea
\end{widetext}

\subsubsection{GP from an Ising spin chain environment}


Let us consider as the environment
a paradigmatic example of quantum criticality:
the Ising spin chain model with a homogeneous
transverse field, with Hamiltonian
\be 
H_{\cal E}(\lambda) = -J \left( \sum_n Z_n
Z_{n+1} + \lambda X_n \right), 
\ee 
where $N$ is the
number of spins in the chain, $J$ the spin-spin coupling, 
$\lambda$ the dimensionless strength of the 
external field, and $X_n$ and $Z_n$ are the Pauli matrices of the n-th spin. 
The quantum critical point is at $\lambda_c=1$ \cite{app:pfeuty}.
If the system spin couples
homogeneously to the Ising chain with strength $\delta$ (i.e. $H_{\cal S E}=\delta
\sum_n Z_n$) this model can be solved analytically using a standard Jordan-Wigner
transformation \cite{app:Quan} into free fermionic modes.
Notice that the requirement of homogeneous coupling is only for simplicity and 
can be relaxed in general \cite{app:Fazio}.
The decoherence factor
induced by this environment ---with the chain initially in the ground state of $H_{\cal E}$--- decomposes 
into a product of factors, each coming independently from a different bath mode \cite{app:Quan},
\be
r(t)
=\prod_{k>0} R_k(t) e^{i (\varphi_k(t)-\varepsilon_k t)}= R e^{i \varphi},
\label{app:IsingDF}
\ee
where
$\varepsilon_k =  2 \sqrt{1+\lambda^2-2 \lambda \cos(k)}$,
with $k=\pm \frac{1}{2}\frac{2 \pi}{N}, \pm \frac{3}{2}\frac{2
\pi}{N},...,\pm \frac{N-1}{2}\frac{2 \pi}{N}$. 
This particular product form  stems from the fact that the 
environment modes are non-interacting, each contributing and independent factor
\ba
R_k(t)&=&\sqrt{\cos^2 \tilde\varepsilon_k t + \sin^2 \tilde\varepsilon_k t \cos^2 2 \alpha_k}, \\
\varphi_k(t)&=&
-i \log \left(
e^{i \varepsilon_k t} \sqrt{
\frac{\cos \tilde\varepsilon_k t + i \sin \tilde\varepsilon_k t \cos 2 \alpha_k}
{\cos \tilde\varepsilon_k t - i \sin \tilde\varepsilon_k t \cos 2 \alpha_k}
}
\right) \nonumber
,\ea where $\tilde\varepsilon_k  =  2 \sqrt{1+(\lambda+\delta)^2-2 (\lambda+\delta) \cos(k)}$, 
$2 \alpha_k=\left[
\theta_k(\lambda+\delta)-\theta_k(\lambda) \right]$, and
$\tan(\theta_k)=\frac{\sin(2\pi k/N)}{\lambda-\cos(2\pi k/N)}.$

\begin{figure}[tbh]
\includegraphics[width=3.5in]{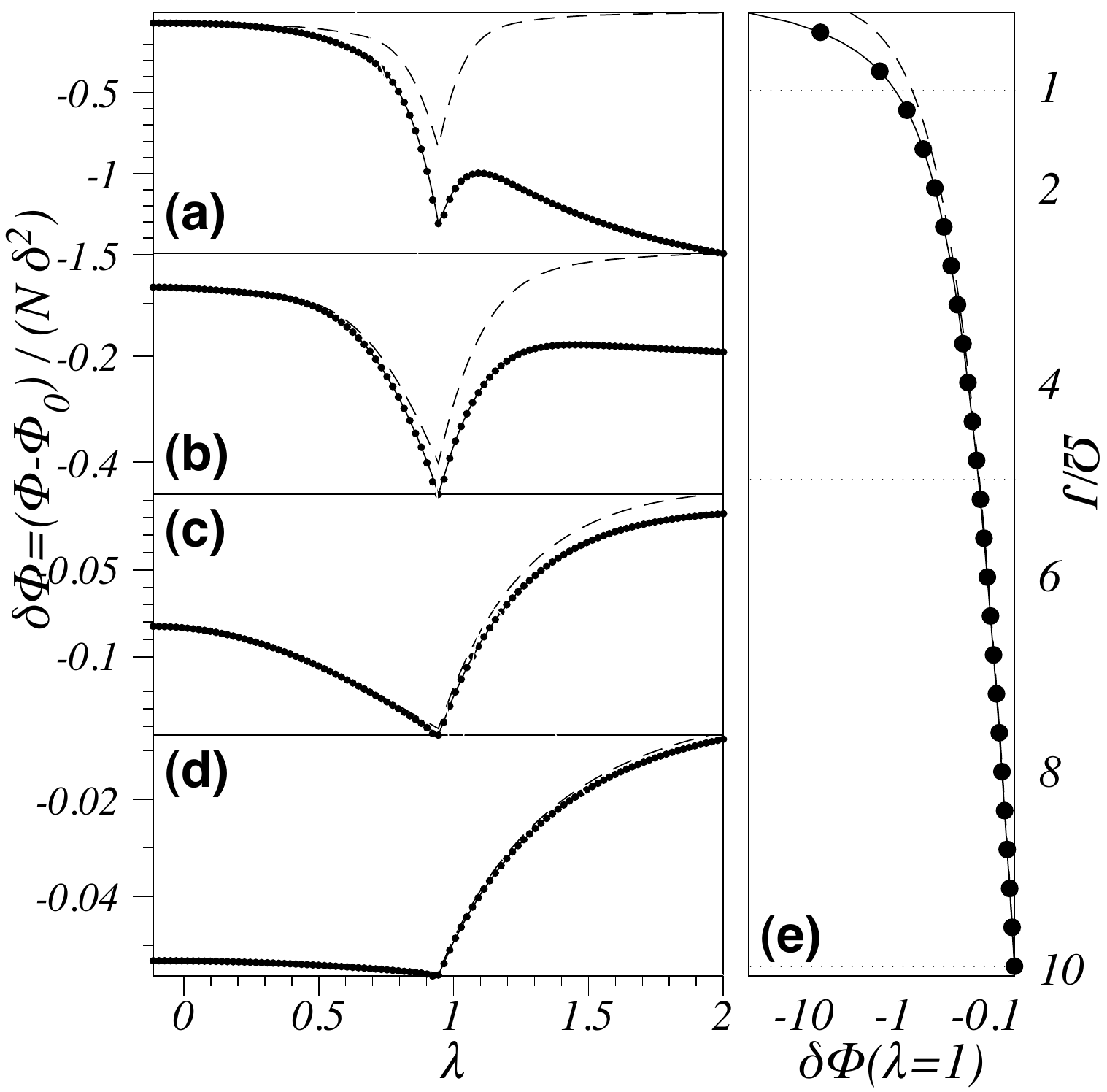}
\caption{
Correction $\delta \Phi=\Phi-\Phi_0$ to the geometric phase of a system spin in presence of an Ising chain 
environment (circles).  In panels {\bf (a)} through {\bf (d)} we show the correction 
as a function of the strength $\lambda$ of the transverse field of the environment chain.
The values of the self--energies of the system are $\Omega=1J, 2J, 5J,$ and $10J$, for 
panels {\bf (a)}, {\bf (b)}, {\bf (c)}, and {\bf (d)} respectively. Here $J$ is the interaction strength
between spins in the environment.
In solid line is the 
third order approximation, and in dashed line the second order one.
The phase correction is shown normalized by the length of the environment chain $N$ and
the strength of the coupling to the environment squared, $\delta^2$. In all plots 
$N=100$ and $\delta=5 \ 10^{-5}J$.
In panel {\bf (e)} we show the correction $\delta \Phi$ at the critical point of the environment ($\lambda=1$),
as a function of the self--energy $\Omega$ of the system. Notice that $\Omega$ is 
inversely proportional to the contact time with the environment, $\tau=2\pi/\Omega$.
The dotted horizontal lines indicate the energies that correspond to the left panels.
}
 \label{app:IsingFig}
\end{figure}

In order to use the approximate expression Eq. (\ref{app:ApproxGP}), we now 
expand each factor $R^2_k(t)$ and $\varphi_k(t)$ in powers of the 
coupling strength $\delta$, 
\ba
R_k^2&=&1-R_{k,(2)}(t)\delta^2-R_{k,(3)}(t)\delta^3+{\cal O}(\delta^4) \nonumber \\
\varphi_k(t) &=& \varepsilon_k t + \varphi_{k,(1)}(t)\delta+{\cal O}(\delta^2),
\ea
and insert them in Eq. (\ref{app:IsingDF}), 
\ba
R^2(t)&=&\prod_{k>0} R_k^2 \simeq \prod_{k>0} \left[1-R_{k,(2)}(t)\delta^2-R_{k,(3)}(t)\delta^3
\right] \nonumber 
\\
&\simeq& 1- \delta^2 \sum_{k>0} R_{k,(2)}(t) - \delta^3 \sum_{k>0} R_{k,(3)}(t)
\nonumber 
\\
&\simeq & 1- \delta^2 \frac{N}{2\pi} \int_0^{\pi }  R_{k,(2)}(t) \, dk - 
  \delta^3  \frac{N}{2\pi} \int_0^{\pi }  R_{k,(3)}(t) \, dk
\nonumber \\
\varphi(t)&=& \delta \sum_{k>0} \varphi_k(t) \simeq \delta  \sum_{k>0}
\varphi_{k,(1)}(t) \nonumber 
\\ &\simeq & \delta \frac{N}{2\pi} 
\int_0^{\pi } \varphi_{k,(1)}(t) \, dk
\ea
where in the last operation we approximate sums over $k$ with an 
integral --- which is a good approximation
in the thermodynamic limit $N\rightarrow \infty$ ---,
and with
\ba
R_{k,(2)}&=&16 \frac{\sin^2 k \sin^2 (\varepsilon_k t) }{\varepsilon_k^4} \nonumber \\
R_{k,(3)}(t) &=& -  \frac{
128 (\cos k-\lambda) \sin^2 k \sin(\varepsilon_k t)
}
{\varepsilon_k ^{6}}
 \nonumber \\
& & \times \left[ \sin(\varepsilon_k t)-\varepsilon_k t \cos(\varepsilon_k t)\right]
\nonumber \\
\varphi_{k,(1)}(t) &=& \frac{ \lambda -\cos k}{\varepsilon_k} 
\ea
%
With these coefficients, the time integrals in Eq.(\ref{app:ApproxGP}) 
can be done analytically using commercial software like Mathematica, which gives us
\ba
\Phi & \simeq & \Phi_0  - \cos \theta \sin^2 \theta \left[
\frac{\delta^2 \Omega}{4} F_2(\lambda)
+ \frac{\delta^3}{24} \left(3 T f_2(\lambda) G_1(\lambda) \right. \right. \nonumber \\
&+& \left. \left. T^3 G^3_1(\lambda)
+ 6  \Omega  F_3(\lambda) -6 G_1(\lambda) F_2(\lambda) \right) \right],
\label{app:approxGPIsing}
\ea
where
\ba
f_2(\lambda)&=&\frac{N}{2\pi}\int_0^\pi
\sin^2 k \frac{\sin(\varepsilon_k T)}{\varepsilon_k^4}, \nonumber \\
F_2(\lambda)&=&\frac{N}{2\pi}\int_0^\pi
\frac{8 T \sin^2 k}
{\varepsilon_k^4}
 \left(1
-\frac{\sin(2 \varepsilon_k T)}{2 \varepsilon_k T}
\right), \nonumber \\
F_3(\lambda)&=&\frac{N}{2\pi}\int_0^\pi
\frac{(\lambda - \cos k)\sin^2 k }{8 \Omega \varepsilon_k^{7}}
\times
\nonumber \\
& &
\left[
4 \pi  \varepsilon_k \left( 2+\cos (2 \varepsilon_k T)
\right)
-3 \Omega \sin (2 \varepsilon_k T)
\right],
\nonumber \\
G_1(\lambda)&=&
\frac{N}{\pi \lambda} \left[(\lambda +1) E\left(\frac{4 \lambda }{(1+\lambda )^2}\right) \right. \nonumber \\
& & \left. +(\lambda-1 ) K\left(\frac{4 \lambda }{(1+\lambda )^2}\right)\right],
\ea
where $E(x)$ and $K(x)$ are the complete elliptic integral and the complete elliptic integral of the first kind
respectively. 
As we can see, the GP of the system spin must be singular at the critical point of the bath
$\lambda=\lambda_c=1$ because
$K(x)$ has a singularity at $x=1$. 

We computed the GP of the system-spin exactly for environment chains of up to $100$ spins.
Longer chains can become computationally unstable and only add fine details to the singularity around
the critical point. We show in Fig.~(\ref{app:IsingFig}) the correction to the GP induced by the coupling to 
the environment as a function of the transverse field of the environment, and for different
cycle periods of the system.
Notice in the figure that the second order approximation to the exact formula performs
poorly compared to the third order one [Eq. (\ref{app:ApproxGP})], especially for large periods $\tau$
where the environment acts for more time.
As is to be expected, 
the duration of contact with the environment also affects strongly the magnitude of the correction to the GP.
At the point where the environment is critical, we observe that the 
GP becomes singular in the thermodynamical limit ---
in contrast to the discontinuity of the GP observed for a first order transition
of the environment \cite{app:Yuang}.
We can see the singularity in the thermodynamical limit appear already in the analytical 
approximate expressions obtained from Eq. (\ref{app:ApproxGP}).

\end{document}